\documentclass{elsart}

\usepackage{graphics}
\usepackage{graphicx}
\usepackage{epsfig}

\usepackage{amssymb}

\begin{document}

\begin{frontmatter}


\title{Hysteresis mediated by a domain wall motion}
\author{Thomas Nattermann}
\address{Institute for Theoretical Physics, University of Cologne,\\
Z\"{u}lpicher Strasse , Cologne, Germany}
\author{Valery Pokrovsky}
\address{
Department of Physics, Texas A\&M University, College Station TX,\\
77843-4242\\
and\\
Landau Institute for Theoretical Physics, Chernogolovka, Moscow\\
District 142432, Russia.}


\begin{abstract}

The position of an interface (domain wall) in a medium with random
pinning defects is not determined unambiguously by a current value
of the driving force even in average. Based on general theory of
the interface motion in a random medium we study this hysteresis,
different possible shapes of domain walls and dynamical phase
transitions between them. Several principal characteristics of the
hysteresis, including the coercive force and the curves of
dynamical phase transitions obey scaling laws and display a
critical behavior in a vicinity of the mobility threshold. At
finite temperature the threshold is smeared and a new range of
thermally activated hysteresis appears. At a finite frequency of
the driving force there exists a range of the non-adiabatic
regime, in which not only the position, but also the average
velocity of the domain wall displays hysteresis.
\end{abstract}

\end{frontmatter}

\section{Introcuction}

This is our tribute to the memory of Per Bak, great scientist and
an exceptional personality. His main contribution to science is
the discovery of a new, very broad class of nonlinear
deterministic systems displaying stable chaotic critical behavior,
which he called Self-Organized Criticality (SOC). It is difficult
to overestimate the significance of this discovery, which has
opened new ways for explanation and unified description of very
remote phenomena, such as earthquakes, plastic deformations,
hysteresis in magnets, dynamics of biosphere and financial
catastrophes. The introduced by Per notion of avalanches is now
central for dynamics of such systems and is most probably a main
source of the $1/f$ noise. Theory of these systems has deserved a
wide popularity and Per is one of the most cited physicists.\\
However, his moral influence in the scientific community was not
less important than his scientific contribution. He \noindent
always was surrounded by people seeking new ways in different
fields: geologists, biologists, physicians, economists, historians
and, certainly physicists. Everybody felt himself free in this
community to criticize everything and to propose new ideas, but
Per's authority and respect to him was extremely high. It was
based on his uncompromising pursuing of truth and hate to any
falsehood in science as well as in social and personal life. Once
he was shown by a big TV company at JFK airport expressing his
indignation by the violation of passenger rights. He was a brave
and noble man. He could be very harsh in discussions, a feature,
which not everybody could tolerate. However, Per was a reliable,
friendly and responsive person ready to help if there was a need
in his help, and for this purpose he could spend his time, money
and influence. We feel that his death has left a vacancy in our
community, which will not be filled.

In this review we consider a simple, but rather wide-spread
phenomenon: the hysteresis mediated by a motion of an interface or
domain wall (DW) driven by an external alternating force in a
medium with random pinning centers. The interplay between the
driving force and elastic and pinning forces develops in time. If
the driving force is constant, it can establish an average
velocity of the DW after a sufficiently long time interval. In the
case of the alternating force this time may be longer or shorter
than period of oscillations resulting in different regimes. The
influence of a heat reservoir gives an additional dimension to
this phenomenon. The interface is a typical system displaying the
SOC\cite{PMB}: in the absence of the external force it exhibits a
fractional structure. A small force results in avalanches on
spatial and temporal scales depending on the force. The stronger
is the force and the higher is the frequency the weaker is the
influence of the avalanches, but they are crucial for the range of
small frequencies and amplitudes important in the
experiment.\\
This problem is a part of a very old hysteresis problem
\cite{steinmetz} occurring in many different dynamical processes:
chemical reactions, magnetization reversal, crystal growth,
absorption and desorption etc., in which many DW are involved. In
this case the interaction between walls must be taken in account.
It is beyond the scope of our consideration, but some rough
estimates can be done on the basis of one-DW theory. On the other
hand, the recent experimental development with its tendency to
sub-micron spatial scale generated many small systems in which the
motion of one or few DW is absolutely realistic. Moreover, the
motion of an individual domain wall was experimentally discovered
and studied by several groups of experimenters \cite{single-exp}.
Therefore, theory of such a motion is necessary for understanding
of the observed phenomena and prediction of new ones.\\
Though the observation of the hysteresis in the ferromagnets
accounts more than 100 years history, the phenomenon could not be
satisfactory explained until the end of 20-th century, when a
theory of DW and its motion in a random medium was developed.
Therefore we start our review with an introductory section briefly
describing this theory. In the next section we consider a DW
moving adiabatically at zero temperature. Adiabatic motion means
that the frequency of the driving force is small enough and the
momentarily velocity is equal to its stationary value at a
constant force equal to the momentarily value of the ac force. The
third section is dedicated to the influence of finite temperature,
but still in adiabatic regime. In the fourth section we consider
non-adiabatic regime predict a new phenomenon: the hysteresis of
velocity (magnetization rate in the case of a ferromagnet). In the
last section we review relevant experiments.

\section{Domain wall in a random medium\label{general}}

\subsection{Zero temperature}
For definiteness we consider a DW in an impure ferromagnet with
uniaxial anisotropy at zero temperature. As it was shown in
\cite{feigel}\cite{koplik}\cite{nat92}\cite{fn93}, equation of
motion for a domain wall without overhangs can be written in a
following way
\begin{equation}
\frac{1}{\gamma}\frac{\partial {\it Z}}{\partial t}= \Gamma
{\bf\nabla}^2 Z + h + \eta ({\bf x},{\it Z}) \label{eq:1}
\end{equation}
where ${\it Z} ({\bf x},t)$ denotes the interface position;
$\gamma$ and $\Gamma$ are the domain wall mobility and stiffness,
respectively; $h$ is the external driving force. For a ferromagnet
$h=\mu_BH{ M}$, where $H$ is the external magnetic field and ${
M}$ is the magnetization. Finally $\eta$ denotes the random force
generated by the impurities. For broad domain walls $\Gamma\approx
J/(a^{D-1}l)$, where $D$ denotes the dimensionality of the wall.
For narrow walls $\Gamma$ depends in general on $J,\;T$, and the
disorder strength in a complicated way \cite{nat82}.\\
The  random forces $\eta ({\bf x},{\it Z})$ generated by pinning
centers obey the Gaussian statistics and have short range
correlations:
\begin{equation}
\overline{\eta ({\bf x},{\it Z})\eta ({\bf x'},{\it Z'})}=\eta^2
l^{D+1} \delta_l({\bf x}-{\bf x'})\Delta(Z-Z'). \label{eq:2}
\end{equation}
Here $\delta_l({\bf x})$ denotes a delta-function smeared out over
a distance $l$. The initial correlator $\Delta_0(Z))$ is an even
analytical function of $Z$, which decays to zero over a finite
distance $l$ and has maximum at $z=0$.\\
In the following we assume that the disorder is weak, i.e. that
$\Gamma\gg\eta l $. Under this assumption the interface is
essentially flat on length scales $L\ll L_p$ (see \cite{nat92}),
where $ L_p\approx l\left(\frac{\Gamma}{\eta
l}\right)^{2/(4-D)}\gg l $ is the so-called Larkin length. On
larger scales the wall adapts to the disorder and gets pinned for
driving fields $h\stackrel{<}{\sim}h_p$ with $ h_p\approx\Gamma
lL_c^{-2}=\eta\left(\frac{\eta l}{\Gamma}\right)^{D/(4-D)} \ll\eta
$ for the pinning threshold. The transverse displacement $w(L)$ of
the DW at the scale $L$ is determined by the roughness exponent
$\zeta$: $w(L)\propto L^{\zeta}$ ($\zeta\leq 1$). The
characteristic time of an avalanche $t(L)$ at the scale $L$ is
determined by the dynamic exponent $z$: $t(L)\propto L^z$.\\
If $h$ exceeds $h_p$, the wall starts to move. Close to the
depinning transition the velocity, which can be considered as an
order parameter of the transition, vanishes according to a power
law
   \begin{equation}
   v\approx v_p\left(\frac {h-h_p}{h_c}\right)^{\beta}, \quad  h>h_p.
   \label{eq:v}
   \end{equation}
Here we introduced the characteristic velocity scale $v_p=\gamma
h_p$. The moving domain wall conserves the rough (fractal)
structure at length scales between $L_p$ and a new scale
established by small velocity
$\xi_v=L_p(v_p/v)^{\frac{1}{z-\zeta}}$, where the pinning force
dominates. At larger scales the bumps on the DW are healed and it
becomes smooth.\\
The functional renormalization group calculations
\cite{nat92}\cite{fn93}\cite{chauve+01} show that $1< z <2$, i.e.
the dynamics close to the depinning transition is {\it
super-diffusive}, reflecting the rapid motion of the object after
the maximum of the {potential} has been overcome. The critical
exponents satisfy new scaling relations \cite{nat92}
   \begin{equation}
   \nu=\frac{1}{2-\zeta}=
   \frac{\beta}{ z-\zeta}\geq \frac{2}{D+\zeta}.
   \label{exponent.relations}
   \end{equation}
These exponents were calculated first to order $\epsilon=(4-D)$ in
\cite{nat92} and recently to order $\epsilon^2$ \cite{chauve+01}:
$\zeta=\frac{\epsilon}{3}(1+0.14331\epsilon)$ and
${z}=2-\frac{2\epsilon}{9}-0.04321\epsilon^2$ \cite{chauve+01}.\\
In the opposite regime $L\gg\xi$ the problem is essentially linear
and $Z$ can be replaced by $vt$ in the argument of $\eta({\bf
x},Z)$. This can be seen qualitatively as follows: On the time
scale $t$ the domain wall advances on average by an amount $vt$.
Randomly distributed pinning centers will lead to a local
distortion which, according to dynamical scaling, spreads over a
region $L(t)\approx L_p\left(\omega_pt\right)^{1/ z}$. The local
retardation or advancement of the object due to the fluctuation in
the density of the pinning centers scales as $ Z(t)\approx
l(L(t)/L_p)^{\zeta}\approx l(\omega_pt)^{\zeta/ z}$, where
$\omega_p=v_p/l$ is the basic frequency at the Larkin length.
Since $\zeta< z$, $ Z(t)$ grows more slowly than $vt$. Thus on
time scales $t>t_{v}=\omega_p^{-1}\left(v_p/v\right)^{ z/(
z-\zeta)}$ and length scales $L>\xi_v\equiv L(t_{v})$ the
non-linearities in the argument of $\eta({\bf x},vt+ Z)$ can be
neglected and the linearized theory applies. In this case $Z({\bf
x},t)$ is replaced by $vt$ in the argument of the random forces.
Random forces act then as a thermal noise with the temperature
$\sim v^{-1}$.\\
After integrating out the interface fluctuations on the length
scales $L\stackrel{<}{\sim}\xi_v$, the effective equation of
motion for the interface profile ${\it Z}({\bf x},t)=\left< {\it
Z} ({\bf x},t)\right>_{\xi_v,t_v} $ where
$\left<\right>_{\xi_v,t_v}$ denotes the spatial and time average
over scales $\xi_v$ and $t_v$, respectively, on large scales
coincides formally with (\ref{eq:1}) with some values substituted
by their renormalized counterparts:  $\gamma\rightarrow\gamma_{\rm
eff}=\gamma\left(\frac{\xi_v}{L_p}\right)^{2-z}$ and $h\rightarrow
h-h_p $. The influence of the random field on these length scales
is negligible .

\subsection{Thermally activated creep of domain walls}\label{thermal}

\begin{figure}[hbt]
   \centerline{\epsfxsize=7cm
   \epsfbox{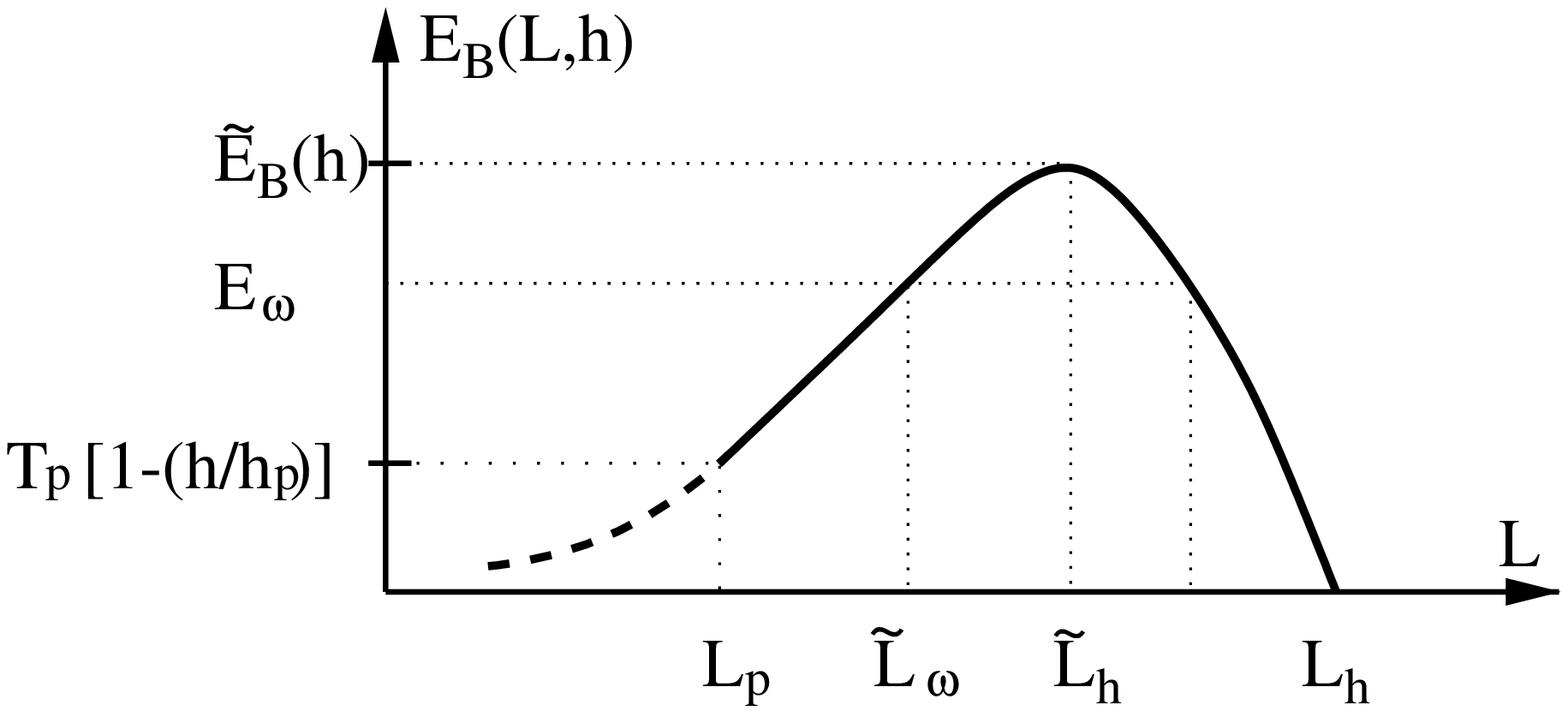}\vspace{-0.7cm}\hspace{2cm}\epsfxsize=5cm\epsfbox{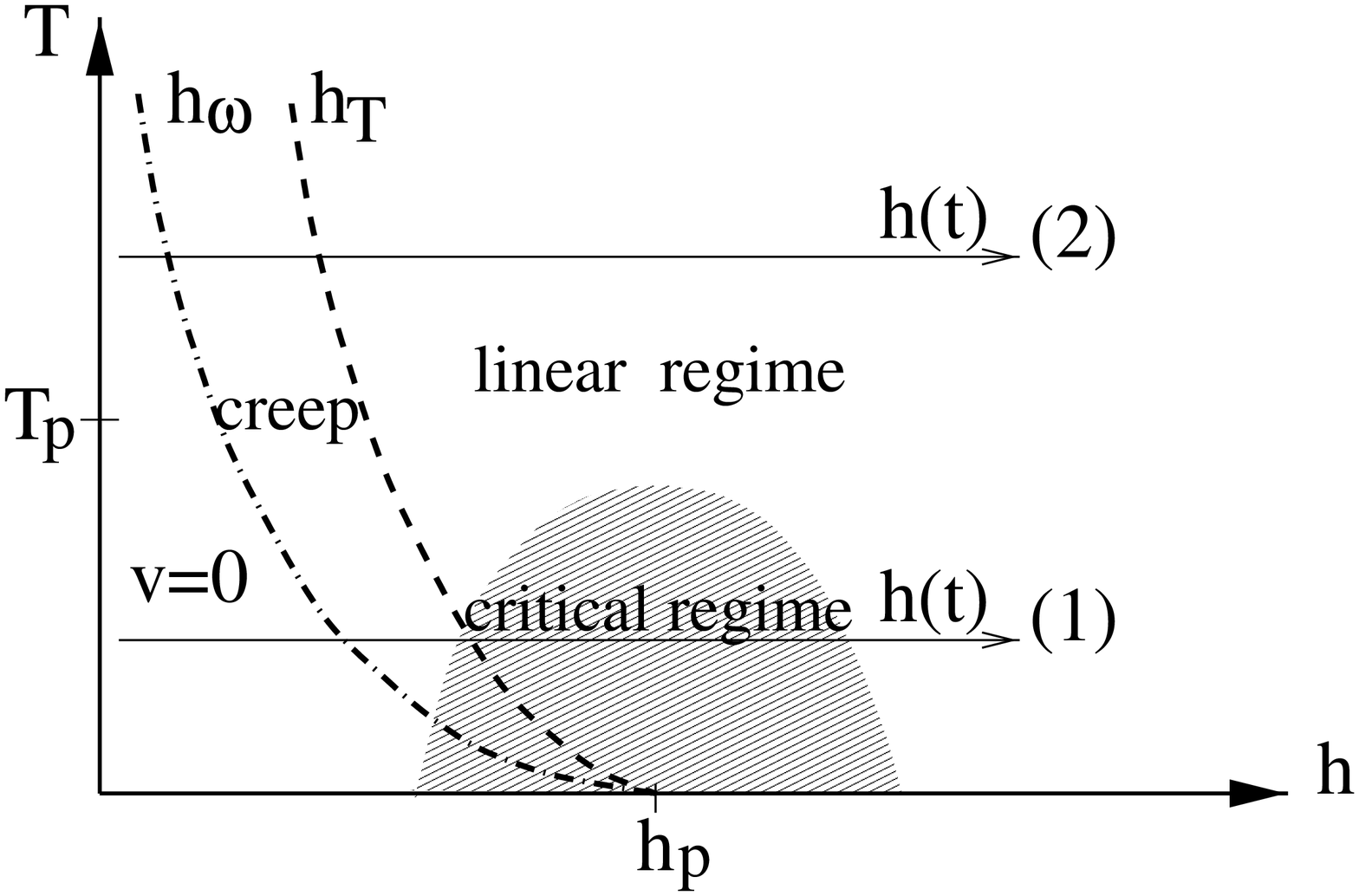}}
\vspace{1cm}
   \caption{\emph{Left}: Energy barrier as a function of
   the length scale $l$ for a given driving force density $h$.
   \emph{Right}: The lines $h_{\omega}$ and $h_T$ as explained in the text.}
   \label{fig:E_B}
   \end{figure}

Next we consider the influence of thermal noise on the interface
motion. Thermal fluctuations lead to a no-zero velocity of the
domain wall even at small forces. All changes in $h$ are still
assumed to be made {\it adiabatically}. The right-hand side of
equation of motion (\ref{eq:1}) must be complemented by {\it
thermal noise} $\zeta({\bf x},{\it Z},t)$ normalized as usual:
\begin{equation}
\langle\eta({\bf x},Z,t)\eta({\bf
x}',Z',t')\rangle_T=2\frac{T}{\gamma} \delta({\bf x}-{\bf
x}')\delta(Z-Z')\delta(t-t')\,. \label{eq:thermal noise}
\end{equation}
In the absence of external driving force the typical free energy
fluctuations of the length scale $L$ are of the order $F(L)=
T_p(L/L_p)^{\tilde\chi}$ where $\tilde\chi=D-2+2\tilde\zeta$. The
{\it energy barriers} between different metastable states scale
have the same order of magnitude. When the driving force $h$
switches on, it changes barriers between neighboring metastable
states at the scale $L$ by the value
$-hw(L)=-hL_p(L/L_p)^{\zeta}$. Thus, the total energy barrier is:
   \begin{equation}
   E_B(L,h)\approx F(L)-hL^Dw_R(L)= T_p\frac{L^{\tilde\chi}}{L_p^{\tilde\chi}}
   \big(1-
\left(\frac{L}{L_h}\right)^{2-\tilde\zeta}\big),
   \label{eq:total_E_B}
   \end{equation}
where we introduced the force length scale
   $
   L_h=L_p\left(\frac{h_p}{h}\right)^{1/(2-\tilde\zeta)}.
   $
A schematic graph of $E_B(L,h)$ vs. $h$ is shown in Figure
\ref{fig:E_B}, it has a maximum at $L=\tilde{L}_h\sim L_h$ and
vanishes for $L=L_h$.
The maximal height of the barrier is:
   \begin{equation}
   \tilde E_B(h)\equiv E_B(\tilde L_h,h)\approx
   T_p\left(\frac{h_p}{h}\right)^{\mu}\,,\quad
   \mu=\frac{\tilde\chi}{2-\tilde\zeta}\,.
   \label{eq:tilde_E_B}
   \end{equation}
It must be overcome by thermally activated hopping to initiate
motion. The {\it creep velocity} of the domain wall follows from
$v_{creep}\approx w(\tilde L_h)/\tau(\tilde L_h)$. According to
the Arrhenius law, the hopping time is $\tau\sim\omega_p^{-1}
e^{\tilde E_B(h)/T}$. Thus, we obtain
   $
   v_{\rm creep}(h)\approx\frac{w(\tilde L_h)}{\tau(\tilde L_h)}\sim
   \exp[{-\frac{T_p}{T}\left(\frac{h_p}{h}\right)^{\mu}}]\,.
  $
We have omitted a prefactor, which is beyond our accuracy. This
formula is valid for $T\ll\tilde E_B(h)$ and was found first by
Ioffe and Vinokur\cite{ioffe.vinokur87} (see also
\cite{Nattermann87}). In the opposite case $T\gg\tilde E_B(h)$ we
expect a linear relation between the driving force and the
velocity: $ v\simeq\gamma h$. The border line between the two
cases i.e.  the inflection point of the curve $v(h)$,
$T\approx\tilde E_B(h)$, defines a {\it temperature dependent
force} $h_T$ (compare Figure 1.)
   \begin{equation}
   h_T=h_p\left(\frac{T_p}{T}\right)^{1/\mu}\,.
   \label{eq:temp_dependent_force}
   \end{equation}
Note that the creep formula is valid only for $h\ll h_T$.\\
Let us now consider the influence of thermal fluctuations on the
depinning transition. At $h\le  h _p$ and $T=0$ the velocity is
zero, but one has to expect that as soon as thermal fluctuations
are switched on, the velocity will become finite. Scaling theory
predicts in this case an Ansatz \cite{Fisher83,Middleton91}
(generalizing (\ref{eq:v}))
   \begin{equation}
   v(h,T)\sim T^{\beta/\tau}\Phi\left(\frac{ h- h _p}{T^{1/\tau}}\right)
   \label{eq:v(f,T)}
   \end{equation}
with $\Phi(x)\to const.$ for $x\to 0$ and $\Phi(x)\sim
x^{\beta/\tau}$ for $x\gg 1$, such that $v( h _p,T)\sim
T^{\beta/\tau}$. $\tau>0$ is a new exponent which still has to be
determined .\\
It is worthwhile to note that relevant thermal fluctuations, which
unpin the domain wall act on scales of the order of $L_p\ll \xi $
as was first indicated by A. Middleton \cite{Middleton91}. At the
critical point $h= h _p$  essentially only barriers on the scale
$L\approx L_p$ are left as we saw earlier. It is therefore
sufficient to consider only this length scale. A detailed analysis
of the form of the effective potential on this scale gives
$\tau=3/2$.

\section{Adiabatic motion of a single domain wall driven by ac force}

\subsection{Dynamics of a rectilinear domain wall at zero
temperature\label{zeroT}}

In this section we describe the motion of a DW, rectilinear at
large scale \cite{LNP}. The reason of the motion is the
alternating driving force, which we assume to have a simple
harmonic shape: $h(t)=h_{0}\sin \omega t$. In previous section we
demonstrated that the DW roughness can be ignored on a time scale
$t>t_{v}$ and the length scale $ L>L_{v}$. Thus, the bending of
the DW can be neglected if \ $\omega t_{v}\ll 1$. Simultaneously
this condition means that the process is adiabatic, i.e. that the
velocity at any moment of time with high precision is equal to its
stationary value corresponding to the value of the driving force
$h(t)$ at the same moment of time. The rectilinear (or plane) DW
can be characterized by one coordinate only. We denote it $Z(t)$.
In adiabatic approximation it satisfies an obvious equation:
\begin{equation}
\frac{dZ}{dt}=\gamma h_{p}f\left( \frac{h(t)-h_{p}}{h_{p}}\right)
, \label{adiabatic}
\end{equation}
where the scaling function $f(x)$ behaves asymptotically as
$x^{\beta}$ at small $x$ and as $x$ at large $x$. Instead of
integrating it over time, we integrate it over field by the
following change of coordinate: $ dt=\frac{dh}{\omega
\sqrt{h_{0}^{2}-h^{2}}}$ Thus, we find a following expression for
$Z$ as function of the field $h$:
\begin{equation}
Z(h,h_{0},Z_{0})=Z_{0}+\frac{\gamma h_{p}}{\omega
}\int\limits_{h_{p}}^{h}f \left( \frac{h^{\prime
}-h_{p}}{h_{p}}\right) \frac{dh^{\prime }}{\sqrt{
h_{0}^{2}-h^{\prime 2}}}  \label{Z(h)}
\end{equation}
This equation is correct for $h>h_{p}$; at smaller positive values
of $h$ the DW does not move and $Z=Z_{0}=const$. Therefore, the
magnetization remains constant until the amplitude of the driving
field reaches the value $ h_{p}$. This value marks the first
dynamical phase transition: the appearance of the hysteresis loop.
Now starting from $Z_{0}=-L$ at $h=0$, where $L$ is the half of
size of a rectangular sample, and applying positive magnetic
field, one can increase $Z$ until it takes its maximum value
\begin{equation}
Z_{\max }=-L+\frac{2\gamma h_{p}}{\omega
}\int\limits_{h_{p}}^{h_{0}}f\left(
\frac{h^{\prime }-h_{p}}{h_{p}}\right) \frac{dh^{\prime }}{\sqrt{%
h_{0}^{2}-h^{\prime 2}}}  \label{Zmax}
\end{equation}
This equation is valid as long as the right-hand side of equation
(\ref{Zmax}) is less than $L$. When it becomes larger, the maximal
value of $Z$ obviously remains equal to $L$. The value $h_{t2}$ of
the amplitude $h_{0}$, at which $Z_{\max }$ becomes equal to $L$
determines the second dynamical phase transition: the complete
reversal of magnetization. The hysteresis loop becomes symmetric
with respect to inversion: $h\rightarrow -h,~M\rightarrow -M$
($Z\rightarrow -Z$) (see Figure 2b). At smaller values of $h_{0}$
between $h_{p}$ and $h_{t1}$ the reversal of the magnetization is
incomplete, the hysteresis loop is symmetric with respect to
reflection $ h\longrightarrow -h$, but not to inversion of
magnetization (see Figure 2a).
\begin{figure}
   \centerline{\epsfxsize=8cm
   \epsfbox{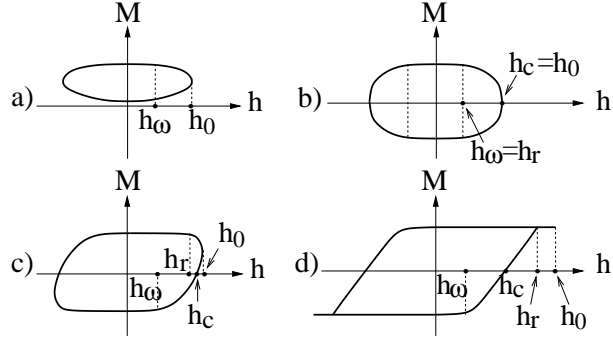}}
   \caption{ Shapes of the hysteresis loops  and dynamical phase transitions}
   \label{loops}
   \end{figure}
Finally, when $Z$ reaches the value $L$ for a quarter of period or
less, i.e. at $h\leq h_{0}$, in a range of fields $h_{r}<h<h_{0}$
the magnetization reaches the saturation (in a little more
realistic model it becomes a single-valued function of the field).
We call these parts of the magnetization graph whiskers (see
Figure 2c). The value $h_{t3}$ of $h_{0}$, at which the whiskers
first appear determines the third dynamical transition. It can be
found from the obvious equation: $ Z(h_{t3},h_{t3},-L)=L$, where
we have used notations of equation (\ref{Z(h)}) . Note that
similar equation is valid for $h_{t2}$: $ Z(h_{t2},h_{t2},-L)=L/2
$ From these equation we find a scaling relation for the phase
transitions fields:
\begin{equation}
\frac{h_{t2}}{h_{p}}=F\left( \frac{\omega L}{2\gamma h_{p}}\right)
;\quad \frac{h_{t3}}{h_{p}}=F\left( \frac{\omega L}{\gamma
h_{p}}\right) , \label{scaling}
\end{equation}
where $F(x)$ is a function inverse to
$G(y)=\int\nolimits_{1}^{y}f(y^{\prime }-1)\frac{dy^{\prime
}}{\sqrt{y^{2}-y^{\prime 2}}}$.\\
It is accepted to characterize the hysteresis loop by the coercive force $%
h_{c}$, i.e. by a value of the driving force at which the
magnetization vanishes ($Z=0$): $Z(h_{c},h_{0},-L)=0$. Earlier we
have introduced the saturation field $h_{r}$ . Equation for it is:
$Z(h_{r},h_{0},-L)=L$. Equation (\ref{Z(h)}) implies that these
two fields divided by $h_{p}$ obey the scaling equations depending
on 2 dimensionless arguments $u=\frac{\omega L}{\gamma h_{p}}$ and
$v=\frac{h_{0}}{h_{p}}$. The reader can find details in the
original article \cite{LNP}. We should mention the asymptotic
behavior at very big $u$ or $v$: $h_{t2}\approx \frac{\omega
L}{2\gamma }$ ; $h_{t3}\approx 2h_{t2}$; $h_{c}\approx
\sqrt{h_{t2}\left( 2h_{0}-h_{t2}\right) }$; $h_{r}\approx
\sqrt{h_{t3}\left( 2h_{0}-h_{t3}\right) }$. The critical
asymptotics near the mobility threshold $\left( v-1\ll 1\right) $
is more complicated. The details can be found in the same article
\cite{LNP}.\\
Numerical MC simulation of the 2-dimensional Ising model with
random bonds and the Glauber dynamics \cite{LNP} demonstrated that
the DWs, initially numerous, quickly merge to a few ones with
linear size of the same order as the size of the sample. This
stage of the magnetization reversal is the longest and determines
dynamics in total. Thus, our results are qualitatively correct for
larger systems at the last and longest stage of the hysteresis
process. The same numerical simulation has reproduced all types of
the hysteresis loop predicted by theory of the single DW
hysteresis. However, the quantitative details may be different.
Moreover, since theory of multidomain hysteresis does not yet
exist, there is no reliable estimate of the time at which only few
domains remain.

\subsection{Motion of a domain wall at finite temperature}\label{finiteT}

As it was shown in subsection \ref{thermal}, the principal
difference of the dynamics at finite temperature is that the DW
can start to move at an arbitrarily weak driving force due to
thermal activation processes. Thus, strictly speaking, there is no
more mobility threshold for the constant driving force. At low
temperature the threshold will be smeared. Besides the finite
temperature establishes new scales of the length $L_T$, time
$t_T\propto L_T^z$, force $h_T$ and activation energy $E_B(T,h)$.
In adiabatic regime it leads to appearance of an effective,
temperature dependent threshold for the ac driving force. In this
section we follow the work \cite{NPV}\\
If the driving force is alternating in time with frequency $\omega
$ and amplitude $h_{0}$, the barriers at the scale $L$, for which
$\omega \tau (L,h)>1$, where $\tau(L,h)=\tau_0\exp{E_B(L,h)/T}$ is
the relaxation time for such a fluctuation, can not be overcome
during one cycle of oscillation. Thus, the global motion of the DW
may be initiated only when the maximum of value $\omega\tau(L,h)$
over $L$ becomes of the order of one. From the condition $\omega
\tau _{\max }(h)=1$ we find a new, frequency and temperature
dependent driving field $h_{\omega }$, which plays the role of
dynamic threshold. It obeys a following equation:
\begin{equation}
\frac{h_{\omega }}{h_{p}}=\left[ \frac{T_{p}}{T\Lambda }\left( 1-\frac{%
h_{\omega }}{h_{p}}\right) ^{1/\mu }\right] ,  \label{dynthresh}
\end{equation}
where $\Lambda =-\ln (\omega \tau _{0})$ (reminder: $\tau
_{0}=\omega_0^{-1}$ is a microscopic hopping time: we assume
$\omega \tau _{0}\ll 1$). At $%
h<h_{\omega }$, there is no macroscopic motion of the wall;,
though still its segments transfer between different metastable
states with avalanches whose development time is less than $2\pi
/\omega$ and the length scale is less than $\tilde
L_{omega}=L_p\left(\frac{T}{T_p} \ln
{\frac{\omega_p}{\omega}}\right)^{1/\tilde\chi}. $. This process
gives rise to dissipation. Drift of the wall as a whole starts at
$h_{0}>h_{\omega }$. At $\omega\tau_0\ll 1$, $h_{\omega }<h_{T}$.
Note that equation (\ref{dynthresh}) determines $h_{\omega }$ as a
monotonously decreasing function of temperature accepting the
value $h_{p}$ at $T=0$. At a fixed temperature $h_{\omega}$ is a
monotonously decreasing function of frequency. Thus, DW subject to
the ac driving force either remains in rest at $h_{0}<h_{\omega
}$, or moves due to thermal activation (creep regime) at
$h_{\omega }<h_{0}<h_{T}$, or it moves in the sliding regime at
$h>h_{T}$. At low temperatures these three regimes overlap with
the critical behavior near $h=h_{p}$. A schematic phase diagram of
the DW motion in variables $T-h$ is shown on the r.h.s of  Figure
\ref{fig:E_B}.\\
Having derived the velocity of DW motion as a function of the
driving field at fixed temperature and frequency, we can follow
the hysteresis curve using the adiabatic equation of motion
similar to that at zero temperature (\ref {adiabatic}), but
containing new parameters $T$ and $h_{\omega }$:
\begin{equation}
\frac{dZ}{dh}\equiv v(h)=\frac{\gamma h}{\omega \sqrt{h_{0}^{2}-h^{2}}}%
f\left( \frac{h}{h_{\omega }},\frac{T}{T_{p}}\right) ,
\label{adiabaticT}
\end{equation}
where $f(x,y)$ is a dimensionless function of dimensionless
arguments which is equal zero at $x<0$ and 1 at $x$ or $y$
infinite.\\
The shape of hysteresis loops at finite temperature is rather
similar to their shape at zero temperature. In particular, all 3
dynamic phase transitions described in subsection \ref{zeroT}
proceed at finite temperature as well, but the role of the
threshold driving force plays the field $h_{\omega }$. At the
first transition the amplitude $h_{0}$ coincides with $h_{\omega
}$. At the second and the third transition amplitudes the same
values are determined by equations:
\begin{equation}
\int\limits_{h_{\omega }}^{h_{tn}}\frac{v(h)dh}{\sqrt{h_{tn}^{2}-h^{2}}}=%
\frac{(n-1)\omega L}{2};~n=2,3
\end{equation}
Equations (\ref{transT},\ref{adiabaticT}) display the scaling
similar to that at zero temperature: the dimensionless ratios
$h_{tn}/h_{\omega }$ are defined by the same function of
dimensionless variable $(n-1)\omega L/(2\gamma h_{\omega })$.  \

\section{Non-adiabatic motion of a wall driven by an ac field}
\subsection{Zero temperature}

We again consider the motion caused by an ac driving force $
h(t)=h_0\sin{(\omega t)}$ in eq. (\ref{eq:1}) with a frequency
$\omega\ll \omega_p=\gamma h_p/l$, but we will not assume
adiabatic condition. We start with zero temperature.
   \begin{figure}[htbp]
   \centerline{\epsfxsize=6cm
   \epsfbox{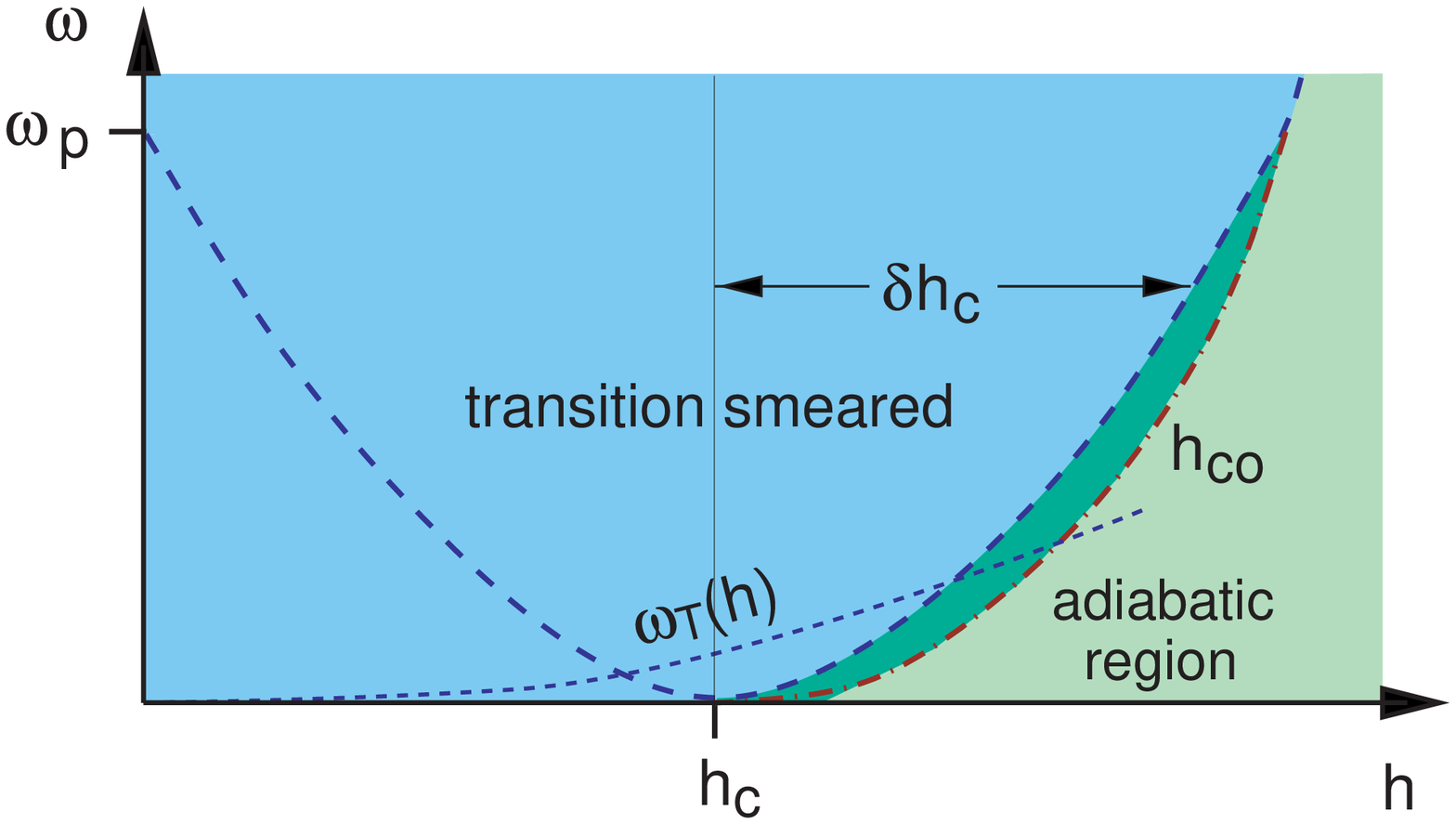}\epsfxsize=6cm\epsfbox{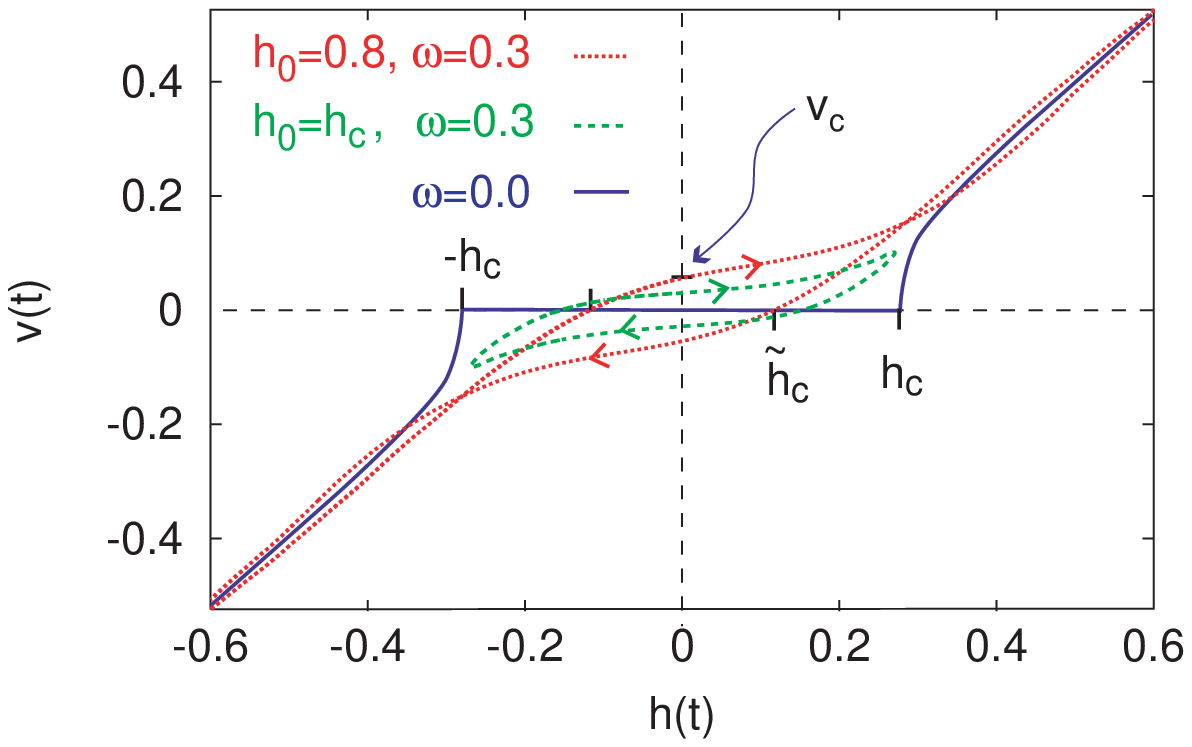}}
   \vspace{0cm}\caption{\emph{Left}: Schematic frequency-field diagram for the depinning in an
   ac external field (with $h_0> h _p$): For $0<\omega\ll \omega_p$ the
   depinning transition is smeared but traces of the $\omega=0$
   transition are seen in the frequency dependency of the velocity at
   $h= h _p$. This feature disappears for $\omega\gg \omega_p$.
   \emph{Right}: Velocity hysteresis of a $D=1$ dimensional interface in
   a random environment.}
   \label{fig:fig2_GNP}
   \end{figure}

The finite frequency $\omega$  of the driving force acts as an
infrared cutoff for the propagation of perturbations generated by
pinning centers. As it follows from (\ref{eq:1})  these
perturbations can propagate during one cycle of the external force
up to the (renormalized) diffusion length $ L_{\omega}=L_p(\gamma
\Gamma/\omega L_p^2)^{1/ z}\equiv L_p(\omega_p/\omega)^{1/ z} $.\\
If $L_{\omega}<L_p$, (i.e. $\omega>\omega_p$) there is no
renormalization and $z$ has to be replaced by 2. During one cycle
of the ac-drive, perturbations generated by local pinning centers
affect the configuration of the domain wall only up to a scale
$L_{\omega}$. If this scale is less than $L_p$, the resulting
curvature force $\Gamma lL_{\omega}^{-2}$ is always larger than
the pinning force and there is no pinning anymore.\\
In the opposite case $L_{\omega}>L_p$ (i.e. $\omega<\omega_p$),
the pinning forces compensate the curvature forces at length
scales larger than $L_p$. As a result of the adaption of the
domain wall to the disorder, pinning forces are renormalized. This
renormalization is truncated at $L_{\omega}$. Contrary to the
adiabatic limit $\omega\rightarrow 0$, there is no sharp depinning
transition at $\omega>0$. Indeed, a necessary condition for the
existence of a sharp transition in the adiabatic case was the
requirement that the fluctuations of the depinning threshold in a
correlated volume of linear size $\xi$, $\delta  h _p\approx  h _p
(L_p/{\xi}_v)^{(D+\zeta)/2}$, are smaller than $( h- h _p)$, i.e.,
$(D+\zeta)\nu\geq 2$ \cite{natter+92}. For $\omega>0$ the
correlated volume has a maximal size $L_{\omega}$ and hence the
fluctuations $\delta  h _p$ are given by
\begin{equation}
   {\delta  h _p}{ }\approx
   h _p\left({L_p}/{L_{\omega}}\right)^{(D+\zeta)/2}=
   h _p\left({\omega}/{\omega_p}\right)^{(D+\zeta)/(2 z)}.
   \label{eq:Delta-h_p}
\end{equation}
Thus, different parts of the domain wall have different depinning
thresholds -- the depinning transition is {\it smeared} in the
interval $\delta h _p$. For a better understanding of the velocity
hysteresis we consider coupling between the different segments of
the domain wall with the average lateral size $L_{\omega}$.
Approaching the depinning transition from sufficiently large
fields, $h_0\gg h _p$ (and $\omega\ll\omega_p$), one observes the
critical behavior of the adiabatic case as long as $\xi_v \ll
L_{\omega}$. The equality $\xi_v\approx L_{\omega}$ defines a
field $h_{co}$ signaling a crossover to an {\it inner} critical
region where singularities are truncated by $L_{\omega}$. Note
that $h_{c0}- h _p= h _p(\omega/\omega_p)^{1/(\nu \tilde
z)}\geq\delta  h _p$ (cf. Figure \ref{fig:fig2_GNP}).\\
A new physical phenomenon occurring in the non-adiabatic regime is
the {\it hysteresis of velocity}. It is shown on the r.h.s. of
Figure \ref{fig:fig2_GNP}, which was obtained by numerical
simulation of one-dimensional version of equation (\ref{eq:1}).
The physical reason for this effect is the hampering of the DW by
the pinning centers. It stops the average motion when the force
becomes small, but not zero. Then the elastic forces tend to
return too stretched intervals of the DW and submit it an average
velocity opposite to the applied force, as is seen on the same
figure. This consideration explains the clockwise circulation on
the central hysteresis loop and appearance of secondary loops. The
reader is referred for details can to reference
\cite{glatz.natter.pokr02}.

\section{Experiments}
In this section we review several experimental works relevant to
our topic. We start with the work by Budde {\it et al.}
\cite{pfnur}. They studied the first order phase transition
between a two-dimensional gas and two-dimensional solid in the
first adsorbed monolayer of a noble gas (Ar, Kr, Xe) on the face
(100) of NaCl. Adsorbed layer was in equilibrium with the 3d gas
of the same atoms. Varying temperature linearly with time back and
forth, they observed the hysteresis of the adsorbate density. The
deviation of the temperature from transition point plays in this
experiment the role of the driving force. They have found that the
width of the hysteresis loop, which corresponds to the coercive
force $h_c$ in magnetic system, scales as $(\Delta T)^{1/4}$,
where $\Delta T$ is the amplitude of the temperature oscillations.
This result agrees with theoretical predictions made in the work
\cite{LNP} (see the end of subsection \ref{zeroT}). This result
suggests that the experimental parameters corresponded to a regime
of the force much larger than the threshold value, so that the
pinning force was negligible.\\
In the works by Kleemann and coworkers \cite{kleemann1,kleemann2}
the authors studied the dielectric spectra in ferroelectric single
crystals ${\rm Sr_{0.61-x}Ba_{0.39}Nb_2O_6Ce_x^{3+}}$ in the
vicinity of its transition temperature. They have found different
behavior of the electric susceptibility $\tilde\chi$ vs.
frequency. At low frequencies they observed a power-like behavior
$\chi\propto\omega^{\gamma}$ with $\gamma$ between 0.2-0.7. Such a
behavior the authors ascribe to the creep motion with a broad
distribution of the activation energy. At higher frequencies they
observed the logarithmic behavior of $\tilde\chi$, which they
treat as the reversible relaxation of the DW segments. The
crossover between these two regimes they attribute to the
dynamical phase transition at $h=h_{\omega}$ as it is predicted in
the work \cite{NPV} (see subsection \ref{finiteT}).

\end{document}